\documentstyle[pasjskel,graphicx,natbib,twocolumn]{article}

\def\commenta{$^*$}
\def\commentb{$^\dagger$}
\def\commentc{$^\ddagger$}
\def\commentd{$^\|$}

\def\an{Astron. Nachr.}

\def\ATsir{Astron. Tsirk.}
\def\ibvs{Inf. Bull. Var. Stars}

\begin{document}

\title{Photometry of the 1991 Superoutburst of EF Pegasi:
       Super-Quasi-Periodic Oscillations with Rapidly Decaying Periods}

\author{Taichi \textsc{Kato}}
\affil{Department of Astronomy, Kyoto University, Sakyo-ku, Kyoto 606-8502}
\email{tkato@kusastro.kyoto-u.ac.jp}


\begin{abstract}
   We observed the 1991 October outburst of EF Peg.  Prominent superhumps
with a period of 0.08705(1) d were observed, qualifying EF Peg as
a being long-period SU UMa-type dwarf nova.  The superhump period showed
a monotonous decrease during the superoutburst, which makes a contrast to
the virtually zero period change observed during the 1997 superoutburst of
the same object.  Large-amplitude, and highly coherent quasi-periodic
oscillations (super-QPOs) were observed on October 18, when superhumps
were still growing in amplitude.  Most strikingly, the QPOs showed a
rapid decrease of the period from 18 m to 6.8 m within the 3.2-hr
observing run.  Such a rapid change in the period has not been observed
in any class of QPOs in cataclysmic variables.  We propose a hypothesis
that the rapid decrease of the QPO period reflects the rapid removal of
the angular momentum from an orbiting blob in the accretion disk, via
the viscosity in a turbulent disk.  A brief comparison is given with
the QPOs in X-ray binaries, some of which are known to show a similar
rapid decrease in the periods.
\end{abstract}


\section{Introduction}

\subsection{Quasi-periodic oscillations}

   Quasi-periodic oscillations (QPOs) are short-period, quasi-periodic
oscillations widely observed in accreting binary systems, such as
cataclysmic variables (CVs) and X-ray binaries
(XBs; see \citet{vanderkli89QPOreview}; more recent
reviews particularly stressing on ``kilohertz QPOs" include
\authorcite{vanderkli99kHzQPOreview}
(\yearcite{vanderkli99kHzQPOreview}, \yearcite{vanderkli00kHzQPOreview}).
There are two major types of ``quasi-periodic" oscillations in
CVs: dwarf nova oscillations (DNOs) and (in a narrower sense) QPOs
(for a review, see \authorcite{war95book} \yearcite{war95book}).

   DNOs are oscillations observed only in dwarf nova outbursts, having
periods 19--29~s, and have long (several tens to $\sim$100 wave numbers)
coherence times (\cite{rob73DNO}; \cite{szk76DNO}; \cite{pat81DNO};
\cite{hil80DNO}).  The amplitudes of DNOs in optical wavelengths are
generally small (usually $<$0.01 mag).  Several models have been proposed
to account for DNOs (and QPOs in general), including vertical or radial
oscillations of the accretion disk \citep{kat78QPO}, reprocessing of
the light by the orbiting blobs \citep{pat79aeaqr}, non-radial pulsations
of the accretion disk (\cite{pap78CVQPO}; \cite{vanhor80DNQPO}),
radial oscillation of the accretion disk (\cite{cox81QPO}; \cite{blu84QPO};
\cite{oku91QPO}; \cite{oku92QPO}), excitation of trapped oscillations
around the discontinuity of physical parameters \citep{yam95DNoscillation},
and oscillation of the boundary layer of the accretion disk
(\authorcite{col98CVQPO} \yearcite{col98CVQPO}, \yearcite{col00CVQPO1}a,b).

   QPOs (other than DNOs\footnote{We use the term ``QPOs" in this sense
in the rest of this paper.}) are more widely seen in CVs, having periods
40~s to several hundred seconds, have much shorter (usually less than
$\sim$10 wave numbers) coherence lengths.  These less coherent QPOs
have phenomenological similarities with QPOs in XBs, both in their
coherence lengths and in the characteristic periods scaled by the
Keplerian timescales at the inner edge of the accretion disk.

   A potentially new class of QPOs (super-QPOs) were discovered during
the 1992 superoutburst of SW UMa \citep{kat92swumasuperQPO}.
These QPOs were only observed during the particular stages (usually the
initial stage) of superoutbursts (for a review of SU UMa-type dwarf novae
and superoutbursts, see \authorcite{war95suuma} \yearcite{war95suuma}),
and have periods of several
hundred seconds, comparable to those of QPOs other than DNOs.  The most
striking features of super-QPOs were the large amplitudes (up to 0.2 mag)
and the long coherence, lasting at least several tens of wave numbers
\citep{kat92swumasuperQPO}.  The super-QPOs observed by
\citet{kat92swumasuperQPO} consisted of sinusoidal components and short,
deep dips.  \citet{kat92swumasuperQPO} also suggested that the QPOs
reported by \citet{rob87swumaQPO} showed only the sinusoidal components,
but sharing all the other characteristics with super-QPOs
observed during the 1992 superoutburst.  \citet{kat92swumasuperQPO}
proposed that an orbiting blob, which incidentally eclipsed the central
part of the accretion disk, was responsible for the large-amplitude
super-QPOs.  \citet{kat92swumasuperQPO} suggested that the reason why
super-QPOs are only observed during the particular stage of superoutburst
may be that such a blob can be formed under the enhanced dissipation
(i.e. turbulent condition) caused by the tidal instability, which is
responsible for superhumps (\cite{whi88tidal}; \cite{hir90SHexcess}).
Studies of super-QPOs have been, however, hindered by the paucity of
examples.  EF Peg is the only other example, which showed the remarkable
time evolution of super-QPOs during the 1991 superoutburst.

\subsection{EF Peg}

   EF Peg was originally discovered as a variable star
\citep{hof35an255407}, who suggested the Mira-type classification.
\citet{esc37efpeg} recorded two additional (bright and faint ones)
maxima, and derived a period period of 157 d in combination with
the maxima recorded by \citet{hof35an255407}.
However, \citet{san50efpeg} only recorded an apparently nonvariable
star of 13.5--14.0 photographic magnitude on 157 plates.
\citet{tse79efpeg} finally revealed that the true EF Peg is a normally
fainter companion to a nonvariable star (labeled as $w$ in 
\cite{tse79efpeg} and \cite{how93efpeg}).  \citet{tse79efpeg} suggested
that EF Peg is a dwarf nova with a large outburst amplitude.
Their observations showed the existence of two types of --- short and long
--- outbursts, which strongly suggested an SU UMa-type dwarf nova
(for a recent review of SU UMa-type stars and their observational
properties, see \cite{war95suuma}).  Upon this information, the
Variable Star Observers League in Japan (VSOLJ) called for a monitoring
campaign for an outburst since 1985.  \citet{ges88efpeg} searched
Sonneberg plates, and found nine outbursts between 1928 and 1986,
confirming the classification by \citet{tse79efpeg}.
No firm outbursts above a magnitude of 12 had been recorded until the
discovery of an outburst by Patrick Schmeer on 1991 October 15.779 UT at
visual magnitude 10.9 \citep{sch91efpegiauc}.  Since then, only two
secure outbursts have been observed in 1995 January and 1997 October,
according to the reports to the VSNET Collaboration\footnote{
  $\langle$http://www.kusastro.kyoto-u.ac.jp/vsnet/$\rangle$
}, and no further outburst has been observed up to 2001 July.
These observations indicate the intrinsically low outburst frequency of
this object.  \citet{pol98TOAD} even suggested that EF Peg is past the
period minimum of the CV evolution.
The first detection of superhumps was made by \citet{kat91efpegiauc}
on 1991 October 18.  More information of this superoutburst is given in
\citet{how93efpeg}.

\section{Observations}

\begin{figure}
  \begin{center}
    \FigureFile(88mm,60mm){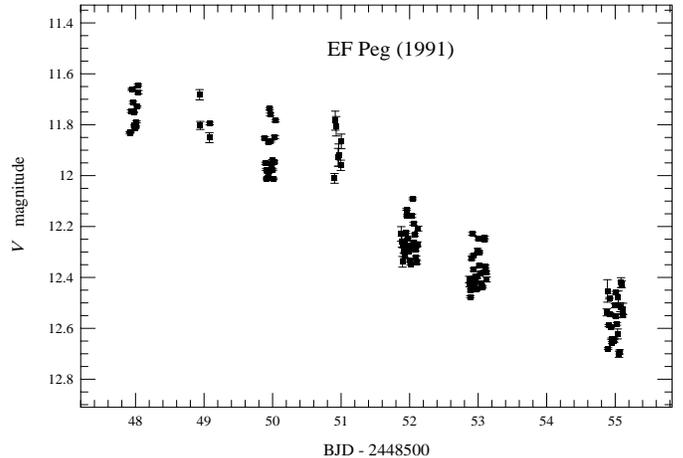}
  \end{center}
  \caption{Overall behavior of the 1991 superoutburst of EF Peg,
  constructed from the CCD observations at Ouda.  Each point and error bar
  represent an average and the standard error of the average of
  observations in each 0.01 d bin.  A constant contribution from the
  $V$=12.71 companion star has been subtracted.
  }
  \label{fig:lc}
\end{figure}

\begin{table}
\caption{Log of observations.}\label{tab:log}
\begin{center}
\begin{tabular}{lrccc}
\hline\hline
UT (start--end)  & N\commenta & mag\commentb & error\commentc & exp\commentd \\
\hline
1991 October     &      &       &       &    \\
18.408 -- 18.542 &  642 & 1.161 & 0.002 & 10 \\
19.429 -- 19.581 &  113 & 1.178 & 0.010 & 10 \\
20.381 -- 20.543 &  793 & 1.277 & 0.002 & 10 \\
21.386 -- 21.551 &  191 & 1.256 & 0.012 & 10 \\
22.373 -- 22.627 & 1161 & 1.494 & 0.002 & 10 \\
23.366 -- 23.625 & 1148 & 1.562 & 0.002 & 10 \\
25.373 -- 25.619 &  741 & 1.673 & 0.003 & 20 \\
\hline
 \multicolumn{5}{l}{\commenta Number of frames.} \\
 \multicolumn{5}{l}{\commentb Magnitude relative to BD +14$^{\circ}$4648;} \\
 \multicolumn{5}{l}{\phantom{\commentb} combined magnitude with the close
 companion.} \\
 \multicolumn{5}{l}{\commentc Standard error of averaged magnitude.} \\
 \multicolumn{5}{l}{\commentd Exposure time (s).} \\
\end{tabular}
\end{center}
\end{table}

   The observations were performed with the CCD camera (Thomson TH7882
chip, 576 $\times$ 384 pixels) attached to the 0.6-m reflector at Ouda
Station, Kyoto University \citep{Ouda} for seven nights between
1991 October 18 and 25.  An on-chip summation of 3$\times$3 pixels
was adopted to minimize the readout time and noise.  A Johnson
$V$-band filter was adopted.  The exposure time was generally 10~s,
except on October 25, when the exposure time of 20~s was used.
The readout time between exposures was 7--9~s.  The total number of
useful frames was 4789.

   A summary of observation is listed in table \ref{tab:log}.
The CCD frames were, after standard de-biasing and flat-fielding,
analyzed using automatic microcomputer-based aperture photometry package
developed by the author.  The diameter of the aperture was 18''.
The magnitudes of the object were determined relatively using a local
standard star BD +14$^{\circ}$4648 = GSC 1117.2097 (star $a$ in
\citet{how93efpeg}: $V$=10.21 and $B-V$=+0.74), whose constancy during the
run was confirmed using a check star GSC 1117.2131 (star $d$ in
\cite{how93efpeg}).  Since EF Peg is a close double (separation
5.3'') with a star (star $w$ in \cite{how93efpeg}), the
resultant aperture photometry is a combined magnitude with this companion.
The $V$-magnitude of the companion was measured as $V$=12.71 on a CCD
from taken on 1992 May 5, when EF Peg was back in quiescence.
As reported in \citet{how93efpeg}, EF Peg was brighter than this companion
during the time of our observation.  No variability has been yet reported
for this companion star.  Barycentric corrections to observed times were
applied before the following analysis.

\section{Results}

\subsection{Outburst light curve}

   The overall light curve of the superoutburst, drawn from the present
observation, is presented in \ref{fig:lc}.  The contribution from the
nearby companion has been subtracted.  The figure corresponds to the
early part of Fig. 2 in \citet{how93efpeg}.  The fading was slower
(0.08 mag d$^{-1}$) for the first three nights, but became more rapid
(0.14 mag d$^{-1}$) after October 20 ($\sim$5 d after the detection of
the outburst).  The decline rate of 0.11 mag d$^{-1}$ in \citet{how93efpeg}
apparently corresponds to an average of these values.  The light curve
of the outburst is a quite typical one for an SU UMa-type superoutburst.

\subsection{Superhump period and profile}

\begin{figure}
  \begin{center}
    \FigureFile(88mm,60mm){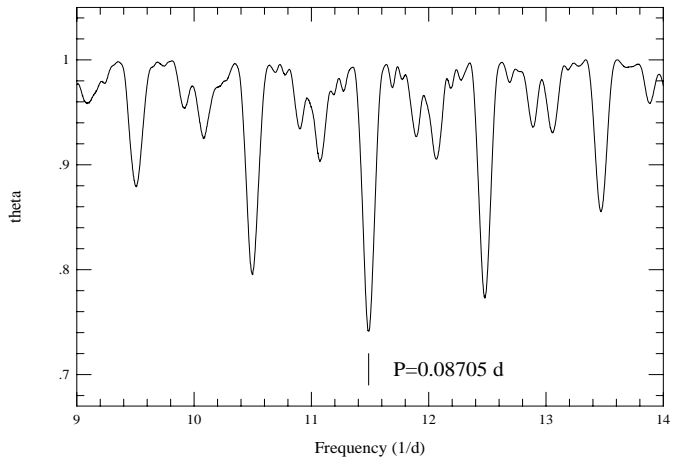}
  \end{center}
  \caption{Period analysis of superhumps in EF Peg.  The Phase Dispersion
  Minimization (PDM) method \citep{PDM} was used, after removing the linear
  trends of decline, and after prewhitening for slow variations with
  frequencies smaller than 2 d$^{-1}$.  The signal at the frequency
  11.488 d$^{-1}$ corresponds to the mean superhump period of 0.08705(1) d.
  }
  \label{fig:pdm}
\end{figure}

\begin{figure}
  \begin{center}
    \FigureFile(88mm,60mm){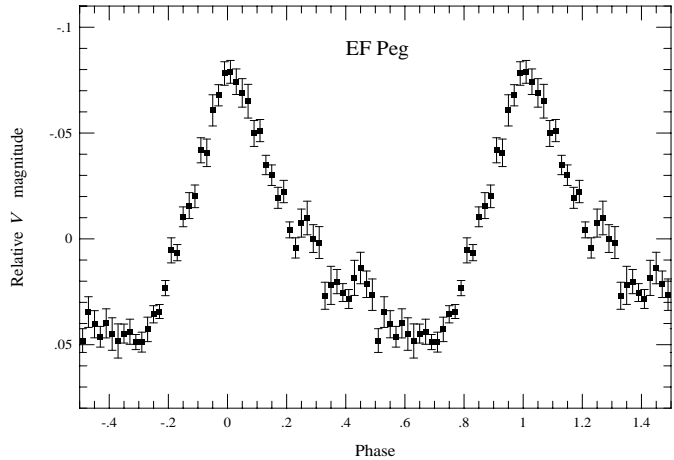}
  \end{center}
  \caption{Phase-averaged light curve of EF Peg superhumps.}
  \label{fig:phaseave}
\end{figure}

   As later shown in \ref{fig:oct18} and also presented by
\citet{how93efpeg}, EF Peg showed strong superhumps during the 1991 October
superoutburst.  In order to determine the superhump period, 
we applied the Phase Dispersion Minimization (PDM) method \citep{PDM}
to all the data, after removing the linear trends of decline, and after
prewhitening for slow variations with frequencies smaller than 2 d$^{-1}$.
The resultant theta diagram is shown in figure \ref{fig:pdm}.
The signal at the frequency 11.488 d$^{-1}$ corresponds to the mean
superhump period ($P_{\rm SH}$) of 0.08705(1) d.  Figure \ref{fig:phaseave}
shows the averaged profile of superhumps.  The phase zero is taken as
BJD (Barycentric Julian Date) 2448547.953 (1991 October 18.453 UT).
The averaged profile of superhumps is a typical one for an SU UMa-type
dwarf nova: a steeper rise and a slower decline \citep{kat98super}.

\begin{figure}
  \begin{center}
    \FigureFile(88mm,120mm){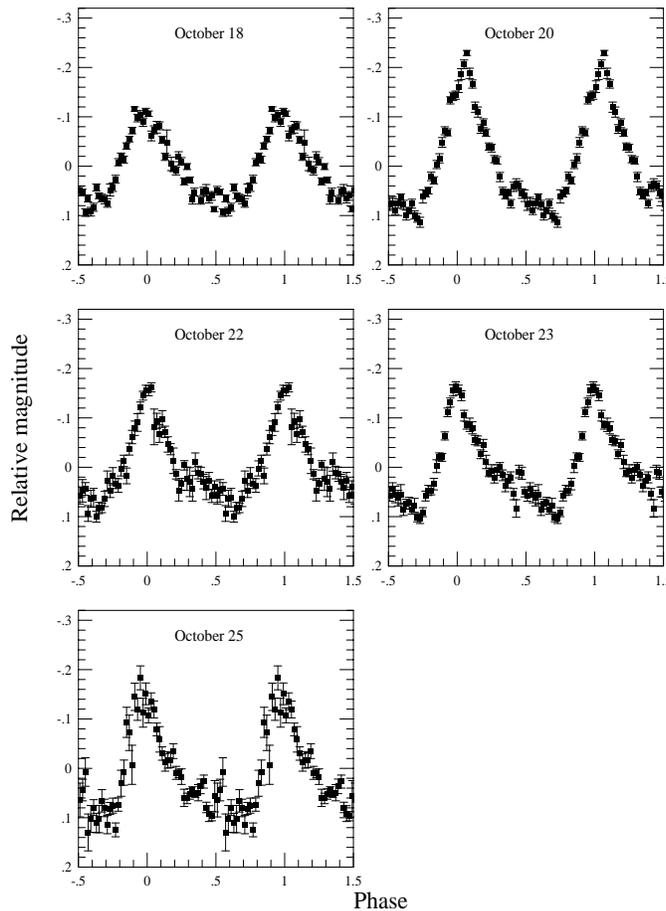}
  \end{center}
  \caption{Evolution of EF Peg superhumps.}
  \label{fig:nightave}
\end{figure}

   Figure \ref{fig:nightave} shows nightly averaged profiles of superhumps.
The amplitude of superhumps reached a maximum on October 20 ($\sim$5 d after
the detection of the outburst), and slowly decayed after that.
The phase zero is defined as BJD 2448547.953.
The shifts of superhump maxima from the zero phases represent the $O-C$
variation described in Section \ref{sec:pdot}.

\subsection{Superhump period change}\label{sec:pdot}

   In order to determine the superhump period change, we extracted by eye
superhump maxima times from our data and from the published light curves
by \citet{how93efpeg}.  The averaged times of a few to several points close
to the maximum were used as representatives of the maxima times.
The errors of maxima times are usually less than $\sim$0.003 d.
The maxima times from the light curves in \citet{how93efpeg} have been
corrected to the common BJD system.  The resultant superhump maxima are
given in table \ref{tab:shmax}.  The values are given to 0.0001 d for the
Ouda data in order to avoid the loss of significant digits in the later
analysis.  The cycle count ($E$) is defined as the cycle number since
BJD 2448547.952 (1991 October 18.452 UT).  A linear regression to the
observed superhump times, gives the following ephemeris.

\begin{equation}
{\rm BJD_{max}} = 2448547.9636 + 0.086868 E. \label{equ:reg1}
\end{equation}

\begin{table}
\caption{Times of superhump maxima}\label{tab:shmax}
\begin{center}
\begin{tabular}{rllc}
\hline\hline
$E$\commenta  & BJD$-$2400000 & $O-C$\commentb & Ref.\commentc \\
\hline
  0 & 48547.9517 & $-$0.0119 & 1 \\
  1 & 48548.0388 & $-$0.0117 & 1 \\
  9 & 48548.750  & \phantom{$-$}0.005  & 2 \\
 23 & 48549.9604 & $-$0.0012 & 1 \\
 46 & 48551.9590 & $-$0.0006 & 1 \\
 47 & 48552.0439 & $-$0.0025 & 1 \\
 52 & 48552.481  & \phantom{$-$}0.000  & 2 \\
 53 & 48552.568  & \phantom{$-$}0.000  & 2 \\
 57 & 48552.9125 & $-$0.0026 & 1 \\
 58 & 48553.0014 & $-$0.0006 & 1 \\
 59 & 48553.0899 & \phantom{$-$}0.0011 & 1 \\
 64 & 48553.543  & \phantom{$-$}0.020  & 2 \\
 80 & 48554.9177 & \phantom{$-$}0.0046 & 1 \\
 81 & 48555.0018 & \phantom{$-$}0.0019 & 1 \\
 82 & 48555.0888 & \phantom{$-$}0.0020 & 1 \\
 87 & 48555.522  & \phantom{$-$}0.001  & 2 \\
 88 & 48555.612  & \phantom{$-$}0.004  & 2 \\
 98 & 48556.480  & \phantom{$-$}0.003  & 2 \\
 99 & 48556.569  & \phantom{$-$}0.005  & 2 \\
110 & 48557.521  & \phantom{$-$}0.002  & 2 \\
111 & 48557.607  & \phantom{$-$}0.001  & 2 \\
122 & 48558.573  & \phantom{$-$}0.012  & 2 \\
133 & 48559.512  & $-$0.005  & 2 \\
145 & 48560.553  & $-$0.007  & 2 \\
156 & 48561.506  & $-$0.009  & 2 \\
157 & 48561.591  & $-$0.011  & 2 \\
\hline
 \multicolumn{4}{l}{\commenta Cycle count since BJD 2448547.952.} \\
 \multicolumn{4}{l}{\commentb $O-C$ calculated against equation
                    \ref{equ:reg1}.} \\
 \multicolumn{4}{l}{\commentc 1: this paper,} \\
 \multicolumn{4}{l}{\phantom{\commentc} 2: \citet{how93efpeg}.} \\
\end{tabular}
\end{center}
\end{table}

\begin{figure}
  \begin{center}
    \FigureFile(88mm,60mm){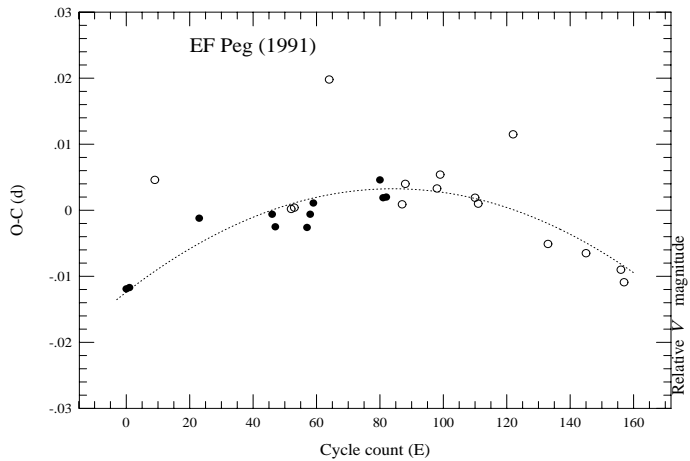}
  \end{center}
  \caption{$O-C$ diagram of superhump maxima.  Filled and open circles
  correspond to maxima times by this work and by \citet{how93efpeg},
  respectively.  The parabolic fit corresponds to equation \ref{equ:reg2}.}
  \label{fig:oc}
\end{figure}

   Figure \ref{fig:oc} shows the $O-C$'s against the linear regression,
equation \ref{equ:reg1}.  The diagram clearly shows the decrease of the
superhump period, which was not apparent in the analysis by
\citet{how93efpeg}, who used nightly determined superhump periods,
rather than the $O-C$ analysis.  After rejecting maxima times
($E$=9, 64, 122) having $|O-C|>0.01$d from a quadratic fit using all
observed times, we obtained the following final quadratic equation.

\begin{eqnarray}
{\rm BJD_{max}} & = 2448547.9512(20) + 0.087240(52) E \nonumber \\
    & - 2.21(31) \times 10^{-6} E^2. \label{equ:reg2}
\end{eqnarray}

   The quadratic term corresponds to $\dot{P}$ = $-$4.4$\pm$0.6 $\times$
10$^{-6}$ d cycle$^{-1}$, or $\dot{P}/P$ = $-$5.1(0.7) $\times$ 10$^{-5}$.

\section{Super-quasi-periodic oscillations}\label{sec:superqpo}

\begin{figure}
  \begin{center}
    \FigureFile(88mm,60mm){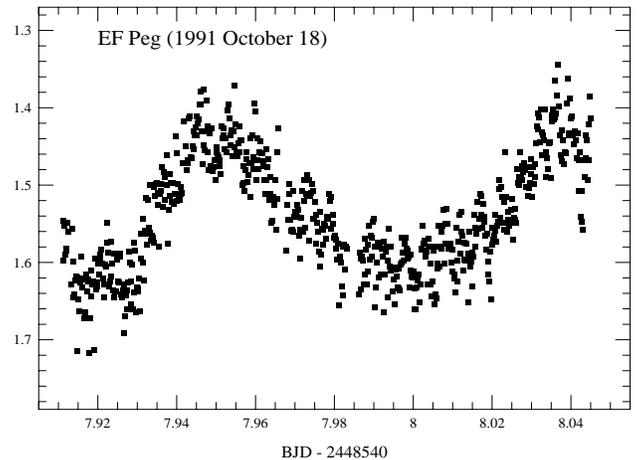}
  \end{center}
  \caption{Light curve on 1991 October 18.  Two superhump maxima
  separated by 0.087 d are clearly seen.  Rapid fluctuations superimposed
  on superhumps represent super-QPOs.
  }
  \label{fig:oct18}
\end{figure}

   The most prominent feature detected during this outburst is
large-amplitude super-quasi-periodic oscillations (super-QPOs:
cf. \cite{kat92swumasuperQPO}), seen on 1991 October 18, $\sim$2 d
before the full growth of superhumps.  These QPOs had disappeared on
October 20.  Figure \ref{fig:oct18} shows the light curve on 1991
October 18.  Two superhump maxima separated by 0.087 d are clearly seen.
The errors of individual
measurements are $\sim$0.01 mag.  The detailed structure in the light
curve therefore reflects the real variation of EF Peg.  Rapid
fluctuations superimposed on superhumps represent super-QPOs.

\begin{figure}
  \begin{center}
    \FigureFile(88mm,60mm){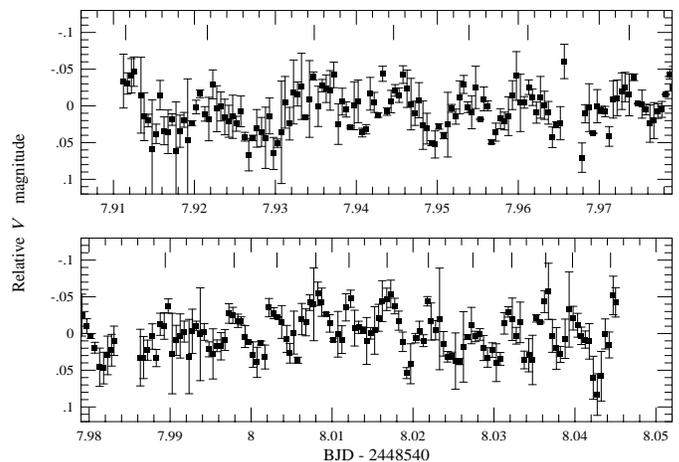}
  \end{center}
  \caption{Super-QPOs, after subtracting the superhump variation from
  \ref{fig:oct18}.  The vertical ticks represent the maxima of super-QPOs
  listed in \ref{tab:qpomax}.  The rapid decrease of the period of
  super-QPOs is clearly seen.
  }
  \label{fig:qpolc}
\end{figure}

   In order to more clearly show the structure of super-QPOs, we have
subtracted superhumps from this light curve, by using a Fourier
decomposition of the superhump profile up to the fourth harmonics.
The resultant residual light curve is shown in figure \ref{fig:qpolc}.

\begin{figure}
  \begin{center}
    \FigureFile(88mm,60mm){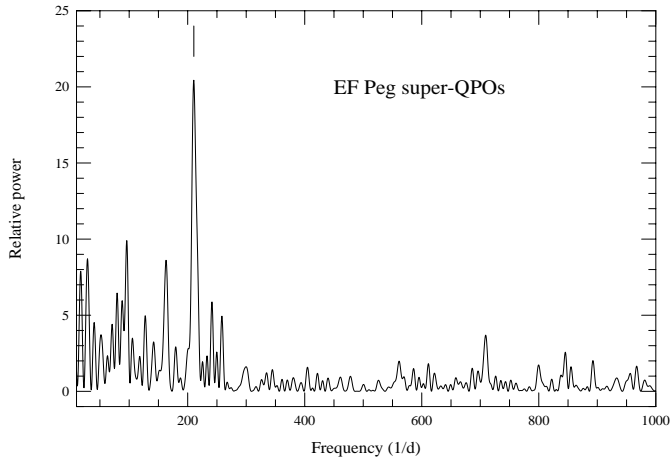}
  \end{center}
  \caption{Power spectrum of super-QPOs.  The strongest signal
  (210 d$^{-1}$) corresponds to the dominant frequency of super-QPOs.
  The presence of considerable power at lower frequencies indicates the
  quasi-periodic nature of the signal.
  }
  \label{fig:qpofou}
\end{figure}

\begin{figure}
  \begin{center}
    \FigureFile(70mm,90mm){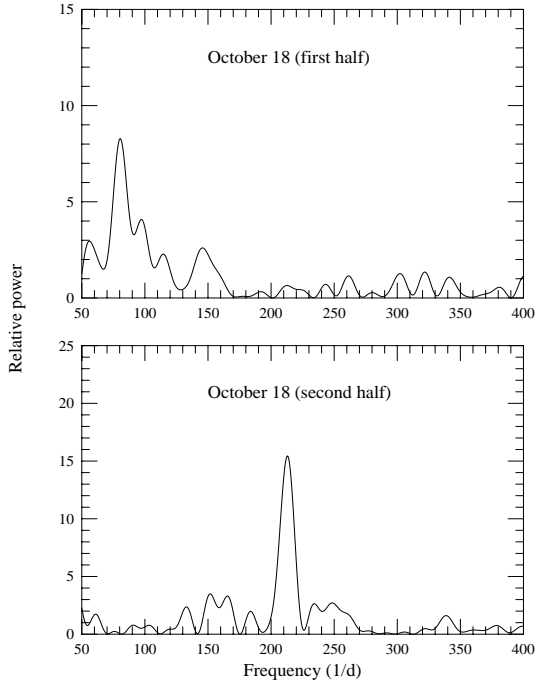}
  \end{center}
  \caption{Variation of the QPO frequency.  The upper and lower panels
  show the power spectra of the first and second half of the October 18
  run, respectively.  The dominant frequencies are 81 d$^{-1}$ (P=18 m)
  and 213 d$^{-1}$ (P=6.8 m), respectively.
  }
  \label{fig:qpovar}
\end{figure}

   Figure \ref{fig:qpofou} shows the overall power spectrum of the
prewhitened data (figure \ref{fig:qpolc}) on October 18.  The strongest
signal (210 d$^{-1}$) corresponds to the dominant frequency of super-QPOs.
The presence of considerable power at lower frequencies indicates the
quasi-periodic nature of the signal.  Figure \ref{fig:qpovar} shows
the variation of the dominant QPO frequency within the October 18 run.
The upper and lower panels show the power spectra of the first and second
half of the October 18 run, respectively.  The dominant frequencies are
81 d$^{-1}$ (P=18 m) and 213 d$^{-1}$ (P=6.8 m), respectively.

   We determined the maxima times of super-QPOs by fitting a parabola
to each super-QPOs (after removing the superhump signal) around their
maxima.  The error of the estimates is typically 0.0002 d.  The resultant
times of maxima are listed in table \ref{tab:qpomax}.  Although the
observation was almost completely continuous, the maxima of $E$=6, 8
fell in short gaps of the observation.

\begin{table}
\caption{Times of maxima of super-QPOs}\label{tab:qpomax}
\begin{center}
\begin{tabular}{rcc}
\hline\hline
$E$\commenta  & BJD\commentb & $O-C$\commentc \\
\hline
 0 & 7.9115 & -0.0130 \\
 1 & 7.9216 & -0.0094 \\
 2 & 7.9348 & -0.0026 \\
 3 & 7.9446 &  0.0007 \\
 4 & 7.9539 &  0.0036 \\
 5 & 7.9612 &  0.0044 \\
 7 & 7.9737 &  0.0040 \\
 9 & 7.9894 &  0.0068 \\
10 & 7.9979 &  0.0088 \\
11 & 8.0032 &  0.0077 \\
12 & 8.0080 &  0.0060 \\
13 & 8.0121 &  0.0037 \\
14 & 8.0168 &  0.0019 \\
15 & 8.0219 &  0.0006 \\
16 & 8.0274 & -0.0004 \\
17 & 8.0322 & -0.0020 \\
18 & 8.0364 & -0.0043 \\
19 & 8.0397 & -0.0074 \\
20 & 8.0444 & -0.0092 \\
\hline
 \multicolumn{3}{l}{\commenta Cycle count since BJD 2448547.9115} \\
 \multicolumn{3}{l}{\commentb BJD$-$2448540.} \\
 \multicolumn{3}{l}{\commentc $O-C$ calculated against equation
                    \ref{equ:regqpo1}.} \\
\end{tabular}
\end{center}
\end{table}

   A linear regression to the observed maxima gives the following equation.

\begin{equation}
{\rm BJD_{max}} = 2448547.9245 + 0.006454 E. \label{equ:regqpo1}
\end{equation}

\begin{figure}
  \begin{center}
    \FigureFile(88mm,60mm){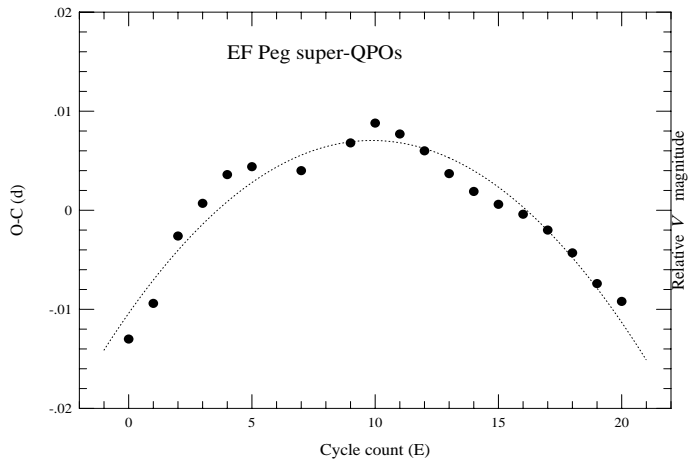}
  \end{center}
  \caption{$O-C$ diagram of maxima times of super-QPOs, listed in
  table \ref{tab:qpomax}.
  The parabolic fit corresponds to equation \ref{equ:regqpo2}.}
  \label{fig:qpooc}
\end{figure}

   The variation of $O-C$'s (figure \ref{fig:qpooc} and equation
\ref{equ:regqpo1} indicates a large contribution of the quadratic
(or higher) term.
The following quadratic equation globally represents the overall $O-C$
variation, although there still exists small residual $O-C$'s to the
quadratic equation.

\begin{eqnarray}
{\rm BJD_{max}} & = 2448547.9141(11) + 0.009988(25) E \nonumber \\
    & - 1.79(12) \times 10^{-4} E^2. \label{equ:regqpo2}
\end{eqnarray}

   The quadratic term corresponds to $\dot{P}$ = $-$2.6$\pm$0.2 $\times$
10$^{-4}$ d cycle$^{-1}$.  By adopting the mean period (0.00645 d) of
super-QPOs from equation \ref{equ:regqpo1}, the period change corresponds to
a decay time-scale of $P/\dot{P}$$\sim$25 cycles, or 0.16 d.

\begin{figure}
  \begin{center}
    \FigureFile(88mm,60mm){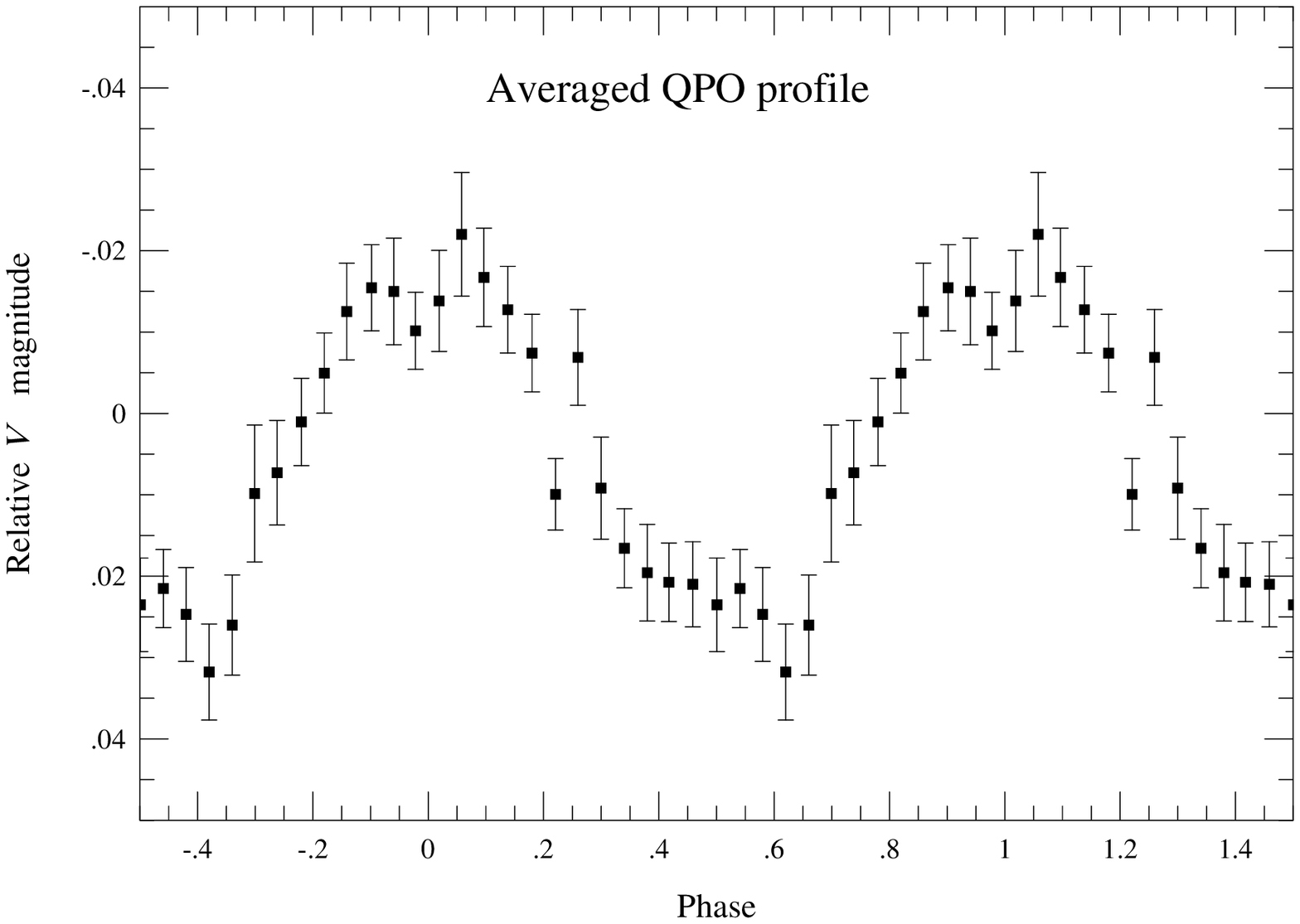}
  \end{center}
  \caption{Averaged profile of the super-QPOs.  The phase zero was
  determined using maxima epochs in table \ref{tab:qpomax} (note that
  cycle lengths of individual QPOs are not constant).  The resultant
  averaged profile has a full amplitude of 0.05 mag.}
  \label{fig:qpoprof}
\end{figure}

   Figure \ref{fig:qpoprof} shows the averaged profile of super-QPOs.
The phase zero was determined using maxima epochs in table \ref{tab:qpomax}.
The resultant averaged profile has a full amplitude of 0.05 mag.

\section{Discussion}

\subsection{Superhump period change}

   As described in Section \ref{sec:pdot}, EF Peg showed a decrease
of the superhump period at a rate of $\dot{P}$ = -4.4$\pm$0.6 $\times$
10$^{-6}$ d cycle$^{-1}$, or $\dot{P}/P$= -5.1(0.7) $\times$ 10$^{-5}$.
This negative period derivative has a typical value for long-period
SU UMa-type dwarf novae (cf. \cite{kat01hvvir}; \cite{pat93vyaqr};
\cite{kat98super}).  However, the same EF Peg showed virtually zero
period change during the 1997 superoutburst (Matsumoto et al.,
in preparation).  While only a few SU UMa-type systems have been observed
during different superoutbursts, all known systems showed relatively
constant $\dot{P}$ even in different superoutbursts \citep{kat01hvvir}.
If such a large variation of $\dot{P}$ in EF Peg is confirmed by
future observations, EF Peg would be a unique system with regards to
the stability of $\dot{P}$.  \citet{kat98super} proposed that positive
$\dot{P}$ in some short-period SU UMa-type systems may be a result
of outward propagation of the eccentricity wave (cf. \cite{lub92SH};
\cite{bab00v1028cyg}).  The large variation of $\dot{P}$, proposed by
the present observation of EF Peg, may reflect the different locations
where tidal instability ignites.

\subsection{Super-QPOs}

   As described in Section \ref{sec:superqpo}, EF Peg showed remarkable
quasi-periodic oscillations (QPOs) during the earliest stage of the
superoutburst, i.e. the growing stage of superhumps (October 18).  These
QPOs had disappeared when the amplitude of superhumps reached the
maximum (October 20).  As seen in \ref{fig:qpooc}, the maximum jump
in the QPO phase was less than 0.002 d, indicating that the oscillations
were essentially coherent during the entire observing run covering 20 QPO
cycles, although a strong global variation of the period was present.
Such a long coherence of oscillations is a signature of super-QPOs
\citep{kat92swumasuperQPO}.  Although the amplitudes of the present
QPOs were smaller than that recorded in SW UMa during the 1992
superoutburst \citep{kat92swumasuperQPO}, they were larger than those of
QPOs observed by \citet{rob87swumaQPO} for the same star.  The observed
QPO profile and amplitude (figure \ref{fig:qpoprof}) resemble those of
the sinusoidal component of \citep{kat92swumasuperQPO}.
From the transient appearance in the early stage of a superoutburst and
the large amplitudes, we consider the present QPOs as a variety of
super-QPOs.  The same explanation in \citet{kat92swumasuperQPO} could
apply to the present large-amplitude QPOs, except that the present QPOs
incidentally lack the dip component.

   The rapid decrease of the QPO period (figure \ref{fig:qpovar},
\ref{fig:qpooc}) is the unique characteristic of the present QPOs.
Such a large variation of the QPO frequency within a short time
(less than 3.2 hr) has never been observed in other QPOs of CVs.
Furthermore, the period decrease occurred continuously (Section
\ref{sec:superqpo}), rather than a sudden mode switch between different
periods.  The phenomenon is hard to explain, if these QPOs arise from
oscillations of the accretion disk, or from the beat between the rotation
of the magnetic white dwarf and the slowly rotating part of the accretion
disk.  Such a rapid decrease of the QPO period can be more naturally
understood if the rotating blob (cf. \cite{kat92swumasuperQPO}), which
rapidly lost its angular momentum via the viscosity in a turbulent disk.
If we assume a 1 M$_\odot$ primary and nearly Keplerian orbits,
a decrease of the period from 18 m to 6.8 m corresponds to a decrease
of the radius of the orbit from 1.6 10$^{10}$ to 8.3 10$^{9}$ cm.
This decrease in 3.2 hr corresponds to a radial velocity of
6.7 10$^{5}$ cm s$^{-1}$ (1/140 -- 1/200 of the orbiting velocity),
which seems to be a reasonable value for accretion during outburst.

\subsection{Comparison with QPOs in XBs}

   Although such a rapid decrease of the QPO period is unique among
CVs, there are a few known similar examples in oscillations
(or large-amplitude QPOs) in some XBs.  The best-known example is seen
in the Rapid Burster (MXB 1730-335).  The first indication of such
oscillations during its type II bursts from the Rapid Burster was found
with the GINGA satellite \citep{taw82RapidBurster}.
\citet{dot90RapidBursterQPO} systematically studied these oscillations,
and found the evidence of strong period changes.
The best example of the ``naked-eye" visibility of large-amplitude
oscillations and the rapid decrease of their periods can be found in
\authorcite{lub92RapidBursteroscillation}
(\yearcite{lub92RapidBursteroscillation})
[see also \authorcite{lub91RapidBursterQPO}
\yearcite{lub91RapidBursterQPO},\yearcite{lub92RapidBurstertype2burst};
for a recent observational review, see also \cite{gue99RapidBurster}).

   The second example of similar oscillations was found in the bursting
X-ray pulsar GRO J1744-28 (\cite{zha96j1744QPO};
\cite{lew96RapidBursterj1744}).  The characteristics of the oscillations
in GRO J1744-28 was studied in detail by \citet{kom97j1744QPO}.
Although the nature of such oscillations are still poorly understood,
several attempts have been made to understand these two unique objects.
In the case of GRO J1744-28, \citet{can96j1744} considered an interplay
between radial and vertical energy transport in the accretion disk,
causing oscillations in the fraction of the gas pressure to the total
pressure.  \citet{can96j1744} showed that these oscillations eventually
lead to a Lightman-Eardley instability \citep{LightmanEardleyInstability},
which \citet{can96j1744} considered to be the cause of Type II bursts.
\citet{can97j1744} further showed that the characteristic QPO frequency
can be reproduced by a selection of parameters.  \citet{abr95XBQPO}
suggested that low-frequency QPOs in some XBs (including those of
the Rapid Burster) and black-hole candidates can be produced via
dissipation of energy in the coronal region above the accretion disk.
While these models were able to reproduce some parts of characteristics
of QPOs (and bursts) observed in these systems, they have not yet
succeeded in explaining a rapid decrease of QPO periods; the cause of this
phenomenon is thus still an open question.
These models for the Rapid Burster and GRO J1744-28 assume the strong
radiation pressure near the inner edge of the accretion disk.  Since
such a condition is hard to achieve in a CV disk, the mechanisms
proposed for QPOs in XBs would not be directly applicable to the
present QPOs in EF Peg.  However, the existence of phenomenologically
similar QPOs both in CVs and XBs may provide future insight for the
fundamental understanding the underlying mechanisms.

\section*{Acknowledgments}

   The author is grateful to Dr. T. Takata and Mr. Y. Tomita for
helping the observation.  The author is also grateful to the VSOLJ and
VSNET observers for supplying vital observations, and especially to
Patrick Schmeer for promptly notifying us of the 1991 outburst detection.

\end{document}